\documentclass[aps,prl,reprint,superscriptaddress]{revtex4-1}

\usepackage{graphicx}
\usepackage{epsfig}
\usepackage{placeins}
\usepackage{amsmath}
\usepackage{nicefrac}
\usepackage{lineno}
\begin{document}

\title{Suppressed decay of a laterally confined persistent spin helix}

\author{P.~Altmann}
\affiliation{IBM Research--Zurich, S\"aumerstrasse 4, 8803 R\"uschlikon, Switzerland}

\author{M.~P.~Walser}
\affiliation{IBM Research--Zurich, S\"aumerstrasse 4, 8803 R\"uschlikon, Switzerland}

\author{C.~Reichl}
\affiliation{Solid State Physics Laboratory, ETH Zurich, 8093 Zurich, Switzerland}

\author{W.~Wegscheider}
\affiliation{Solid State Physics Laboratory, ETH Zurich, 8093 Zurich, Switzerland}

\author{G.~Salis}
\email{gsa@zurich.ibm.com}
\affiliation{IBM Research--Zurich, S\"aumerstrasse 4, 8803 R\"uschlikon, Switzerland}

\begin{abstract}
We experimentally investigate the dynamics of a persistent spin helix in etched GaAs wire structures of 2 to 80~$\mu$m width. Using magneto-optical Kerr rotation with high spatial resolution, we determine the lifetime of the spin helix. A few nanoseconds after locally injecting spin polarization into the wire, the polarization is strongly enhanced as compared to the two-dimensional case. This is mostly attributed to a transition to one-dimensional diffusion, strongly suppressing diffusive dilution of spin polarization. The intrinsic lifetime of the helical mode is only weakly increased, which indicates that the channel confinement can only partially suppress the cubic Dresselhaus spin-orbit interaction.
\end{abstract}

\maketitle

\begin{figure}
\includegraphics[width=0.5\textwidth]{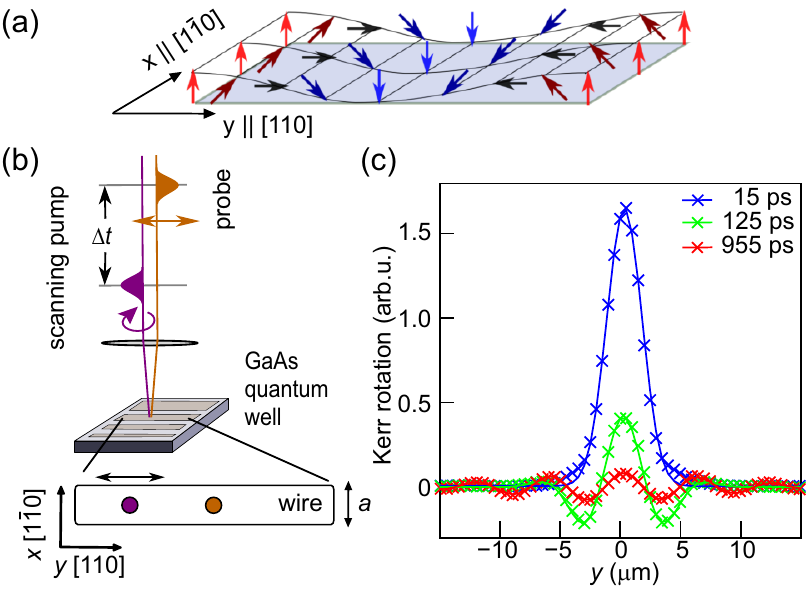} %width x2 in reprint
\caption{\label{fig:fig1}(a) Schematic sketch of the persistent spin helix mode. (b) Measurement principle. A pump and a probe laser pulse are focused on the sample. Their time delay can be tuned, and the position of the pump spot can be scanned relative to the probe position. (c) Scans of time-resolved Kerr rotation measured along $y$ on the 2-$\mu$m-wide wire (crosses) for 3 different time delays. The oscillations visible at $t=125$ and 955\,ps show the formation of a spin helix mode. Lines are fits with a Gaussian times a cosine function.}
\end{figure}

In the quest to enable spintronic technology, it remains a fundamental task to keep the electron-spin lifetime high in systems with spin-orbit interaction (SOI). While effective magnetic fields arising from the SOI can be used to manipulate spins, they also lead to spin decay. Because SOI depends on the electron momentum, electron scattering causes fluctuations of those magnetic fields, giving rise to the so-called D'yakonov-Perel spin dephasing mechanism~\cite{Dyakonov1972}. Such spin decay can be suppressed when the SOI is designed to a symmetry point where spatial position and spin precession angle are correlated. The eigenmode of such a system is the so-called persistent spin helix (PSH) \cite{Schliemann2003, Bernevig2006}, where spins collectively precess over 2$\pi$ within the spin-orbit length, $l_\textrm{SO}$ [see Fig.~\ref{fig:fig1}(a)]. The PSH has been considered for the realization of a diffusive spin-transistor~\cite{Schliemann2003, Kunihashi2012APL}. Although the lifetime of this helical mode is experimentally found to be strongly enhanced~\cite{Koralek2009, Walser2012, Ishihara2014Gate}, it is still limited by imperfect balancing of the SOI, by the cubic Dresselhaus term and possibly by other spin-decay mechanisms~\cite{Lueffe2011, Liu2012, Lueffe2013}.

Additionally it has been proposed to laterally confine the 2D electron gas in a wire structure. For $a< 1/k_\textrm{F}$, where $k_\textrm{F}$ is the Fermi wavevector and $a$ the wire width, the system is purely 1D because of quantization of the momentum transverse to the channel~\cite{Schliemann2003, Pershin2004, Quay2010}. The effective magnetic field has the same magnitude but different signs for the two possible values of the electron momentum along the channel, $\pm \hbar k_\textrm{F}$ ($\hbar$ is the reduced Planck constant). This leads to spin eigenmodes of helical nature that are protected against spin decay. Such systems were first considered for spin transistors~\cite{Datta1990} and are an important building block for the realization of Majorana Fermions, which could potentially be used for quantum computing~\cite{Lutchyn2010, Oreg2010, Alicea2010, Mourik2012}.
In the quasi-ballistic case $1/k_\textrm{F} < a < l_\textrm{p}$, where $l_\textrm{p}$ is the electron mean free path \footnote{Note that $l_\textrm{p}$ is strongly affected by electron-electron scattering \cite{Weber2005} and is, therefore, much shorter than for electron transport. }, the spin dynamics is approximately the same as in the 1D case, if scattering at the edge is specular and in the limit of $a\ll l_\textrm{SO}$, where $l_\textrm{SO}$ is the spin-orbit length~\cite{Chang2009}. 
Surprisingly, even in the diffusive limit $a > l_\textrm{p}$, an enhancement of the spin lifetime was found in optical~\cite{Holleitner2006, Denega2010, Ishihara2013}, transport~\cite{Schaepers2006, Lehnen2007, Kunihashi2009} and theoretical studies~\cite{Malshukov2000, Kiselev2000, Chang2009, Liu2010, Zarbo2010, Kunihashi2012} for wires with $a<l_\textrm{SO}$. 
Numerical~\cite{Kiselev2000, Chang2009} and analytical~\cite{Malshukov2000, Chang2009} studies predict a decay rate of the spin eigenmode that scales with $(a/l_\textrm{SO})^2$. Optical experiments in wires have, so far, mainly used laser spots that are much larger than the helix period. Hence, an averaged spin lifetime was measured without resolving the lifetime $\tau_{\textrm{PSH}}$ of the helical mode. A first measurement of a PSH in wires was presented in Ref.~\onlinecite{Ishihara2014}, but no systematic study on the impact of the wire width on diffusion and on the spin lifetime was conducted.

In this study, we determine the intrinsic lifetime of the helical mode in a PSH system and systematically study the effects of lateral confinement for variable wire width.
We perform optical measurements that track the spatially resolved spin polarization after exciting electron spins locally in a geometry in which wires are patterned along the direction of the PSH. We find that wire confinement strongly suppresses the decay of a PSH, mainly caused by a reduction of the diffusive spin decay when the width of the spin distribution becomes comparable to $a$. Additionally, a small increase of $\tau_\textrm{PSH}$ is found, which we interpret as an implication of the cubic Dresselhaus term that limits the spin lifetime already in the 2D case. Our findings provide deeper insight into previous studies on spin dynamics in wires, and help design spin diffusion channels with reduced spin decay, also away from the PSH regime.

A modulation-doped 12-nm-thick GaAs/AlGaAs quantum well was grown by molecular beam epitaxy on a (001) GaAs substrate. The electron sheet density is $n_\textrm{s} = 3.7 \times 10^{15}$ m$^{-2}$. The quantum well is designed such that the Rashba SOI (characterized by the parameter $\alpha$) and the Dresselhaus SOI (characterized by the parameter $\beta=\beta_1-\beta_3$, with $\beta_1$ the linear, and $\beta_3$ the cubic contribution) are of similar strength. The existence of a PSH in this wafer was reported in Ref.~\onlinecite{Walser2012, Salis2014}. Because $\alpha\approx +\beta$, the PSH direction is along the [110] crystallographic axis, which we define as the $y$-axis. Wire stripes along $y$ were fabricated by photolithography and wet-chemical etching with their nominal widths $a$ ranging from 2 to 80~$\mu$m. The individual wires are separated by
20 $\mu$m spacers.

\begin{figure}
\includegraphics[width=0.5\textwidth]{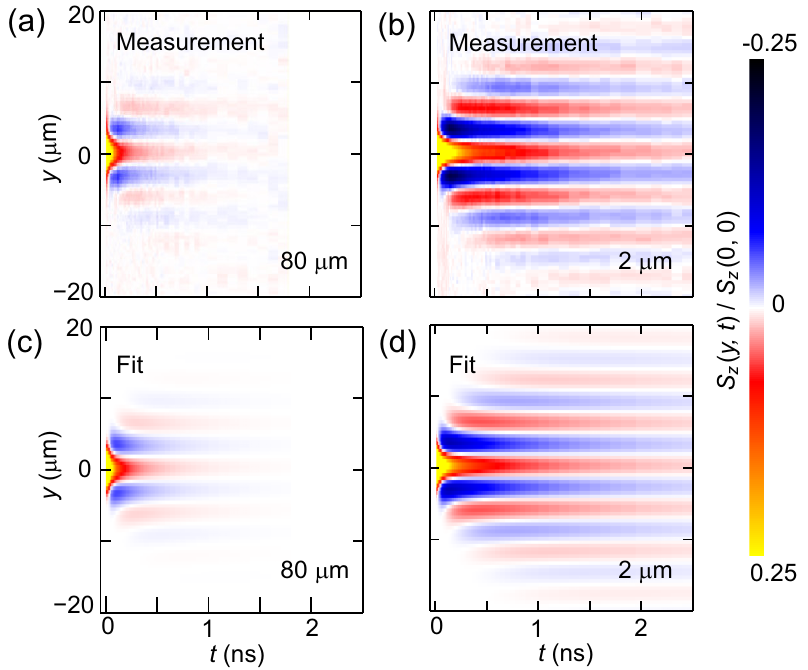} % width x2 in reprint
\caption{\label{fig:fig2} Measurements and fits of $S_{z} (y,t)$. (a) and (b) show 2D plots of the line scans against the time delay between pump and probe pulse. The widest wire, 80~$\mu$m, is shown in (a) and the thinnest, 2~$\mu$m, in (b). The color scale codes the out-of-plane spin polarization $S_{z} (y,t)$ normalized to its maximum value $S_{z} (0, 0)$. (c) and (d) show corresponding fits with the model of Eq.~(\ref{eq1}).}
\end{figure}

For the investigation of the spin dynamics, we use scanning Kerr microscopy with highly focused Gaussian spots with a sigma width $w \approx$~1.5~$\mu$m. Fig.~\ref{fig:fig1}(b) shows a schematic sketch of the measurement principle. Two laser pulses are used in a pump-probe technique. The pump pulse is circularly polarized and excites electron spins with their orientation perpendicular to the QW plane because of the optical orientation effect \cite{MeierBook}. With a defined time delay, a probe pulse arrives that is linearly polarized. The polarization axis of the reflected probe pulse is rotated because of the magneto-optical Kerr effect by an angle proportional to the out-of-plane spin polarization at the location of the spot. By scanning the relative distance between pump and probe spots, a spatial spin distribution can be mapped out at varying time delays. All measurements in this report have been performed at a sample temperature of 40~K.

Figure~\ref{fig:fig1}(c) shows Kerr-rotation measurements as a function of the relative distance $y$ along the wire between the pump and probe spots and for different time delays, $t$, recorded at the transverse center position $x=0$ on the 2-$\mu$m wire. The Kerr rotation angle is proportional to the out-of-plane spin polarization, $S_{z} (y)$, and at $t=15$~ps, resembles the Gaussian intensity profile of the pump spot, although broadened by diffusion. The measurement at $t=955$~ps exhibits a cosine oscillation of $S_{z}(y)$, a direct signature of the spin helix that forms with a wave number $q_0 = 2 \pi / \lambda_{\textrm{PSH}}$ (helix period $\lambda_\textrm{PSH}$). For all times, the data can be very well fitted [solid lines in Fig. \ref{fig:fig1}(b)] by a product of a cosine and a Gaussian that broadens in time owing to diffusion (quantified by the diffusion constant $D_\textrm{s}$).

Figures~\ref{fig:fig2}(a) and \ref{fig:fig2}(b) show color-scale plots of the temporal evolution of line scans along wires of 80~$\mu$m and 2~$\mu$m width. Both, diffusive broadening as well as spin helix formation can be seen. Whereas for both wires the signal decays with time, the visibility of the PSH in the 2-$\mu$m-wide wire is largely enhanced at longer times.

To distinguish between an enhancement of the lifetime $\tau_\textrm{PSH}$ of the helical eigenmode and a transition to 1D diffusion, we fit the data with a model for the dynamics of a localized spin excitation for a system close to the PSH regime \cite{Salis2014}:

\begin{multline}
\label{eq1}
S_{z} (y, t) = S_{z} (0, 0) \cdot \cos \left(  q_0 y \cdot \frac{  2 D_\textrm{s} t}{w^2 + 2 D_\textrm{s} t} \right) 	\\
\cdot \exp \left( - \frac{y^2}{2 (  w^{2} + 2 D_\textrm{s} t)}  \right)
\cdot A_{\textrm{diff}} (t) \cdot A_{\textrm{add}} (t) \cdot A_{\textrm{dec}} (t) ,
\end{multline}
with
\begin{eqnarray}
\label{eq2}
A_{\textrm{diff}} (t) = \left( \frac{ w}{\sqrt{   w^2 + 2 D_\textrm{s} t}} \right)^{dim} , \\% q'_0 = \left( 1 - 
\label{eq3}
A_{\textrm{add}} (t) = \exp \left( - D_\textrm{s} q_0^2 \frac{w^2}{w^2 + 2 D_\textrm{s} t}t \right),\\
\label{eq4}
A_{\textrm{dec}} (t) = \exp \left( - \frac{t}{\tau_\textrm{PSH}} \right).
\end{eqnarray}

In addition to $A_{\textrm{dec}} (t)$, describing the exponential decay of the PSH eigenmode, the terms $A_{\textrm{add}} (t)$ and $A_{\textrm{diff}} (t)$ contribute to the signal decay. $A_{\textrm{add}} (t)$ arises because the helical mode evolves out of a spatially localized spin density of finite width, $w$ \cite{Salis2014}. For $w \rightarrow 0$, $A_\textrm{add} \rightarrow 1$.

The diffusion factor, $A_{\textrm{diff}} (t)$, accounts for the diffusive dilution of the spin density as the radius of the distribution increases in time with $\sqrt{2 D_\textrm{s} t}$. For 2D diffusion ($dim = 2$), it is proportional to $1 / 2 D_\textrm{s} t$. In the 1D limit ($dim = 1$), the spins can only diffuse along one direction, and accordingly the signal will decay only with $1/\sqrt{2 D_\textrm{s} t}$.

\begin{figure}
\includegraphics[width=0.5\textwidth]{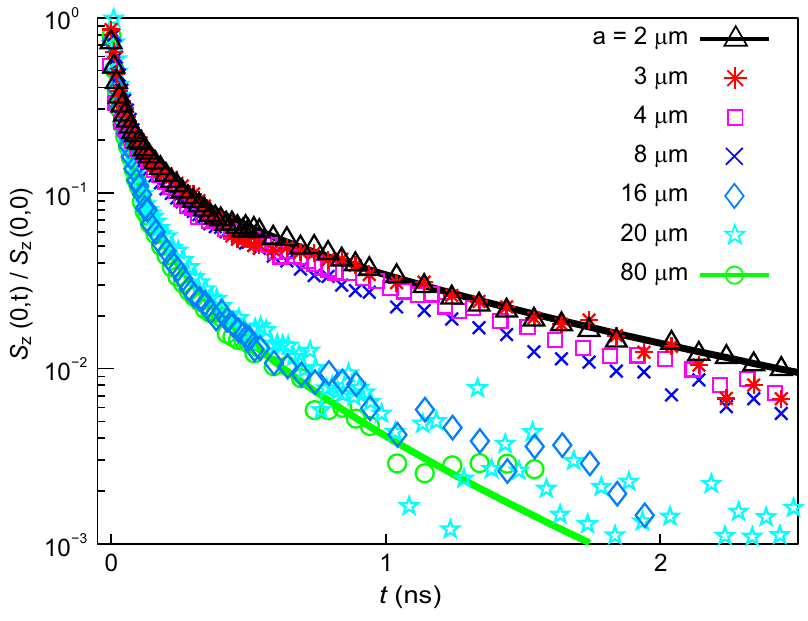}
\caption{\label{fig:fig3} Decay of $S_{z} (0,t)$ for different wire widths $a$. Symbols show the amplitude of individual fits to line scans [see Fig.~\ref{fig:fig1}(b)], whereas the solid lines exemplarily show global fits according to Eq.~(\ref{eq1}) for the 80-$\mu$m and the 2-$\mu$m wires.}
\end{figure}

We fit the data $S_{z} (y, t)$ to the model of Eq.~(\ref{eq1}) and show the results for the 80-$\mu$m and 2-$\mu$m wires in Figs.~\ref{fig:fig2}(c) and (d). The model reproduces the measurements very well. Fit parameters are the initial distribution width $w$, the PSH wave number $q_0$, the diffusion constant $D_\textrm{s}$, the maximum amplitude $S_{z} (0, 0)$, the PSH lifetime $\tau_\textrm{PSH}$, and the dimensionality factor $dim$.
Figure~\ref{fig:fig3} shows the temporal evolution of the normalized amplitude, $S_{z} (0, t) / S_{z} (0, 0)$, for all wire widths. The solid lines are obtained from fits of the full data set, $S_{z} (y,t)$, with Eq.~(\ref{eq1}). For better readability, we only show the results for the widest and the narrowest wire. The symbols are the results obtained from fitting individual line scans, as depicted in Fig. \ref{fig:fig1}(b). As the results of both fitting methods coincide very well for all wires, we conclude that the model of Eq.~(\ref{eq1}) captures all relevant decay mechanisms. For all curves, the decay in the first few hundred picoseconds is dominated by the additional decay terms, Eqs.~(\ref{eq2}) and (\ref{eq3}). Two groups of wires can be distinguished: In wires narrower than 8~$\mu$m, the Kerr signal decays significantly slower than in the other wires.

\begin{figure}[b]
\includegraphics[width=0.5\textwidth]{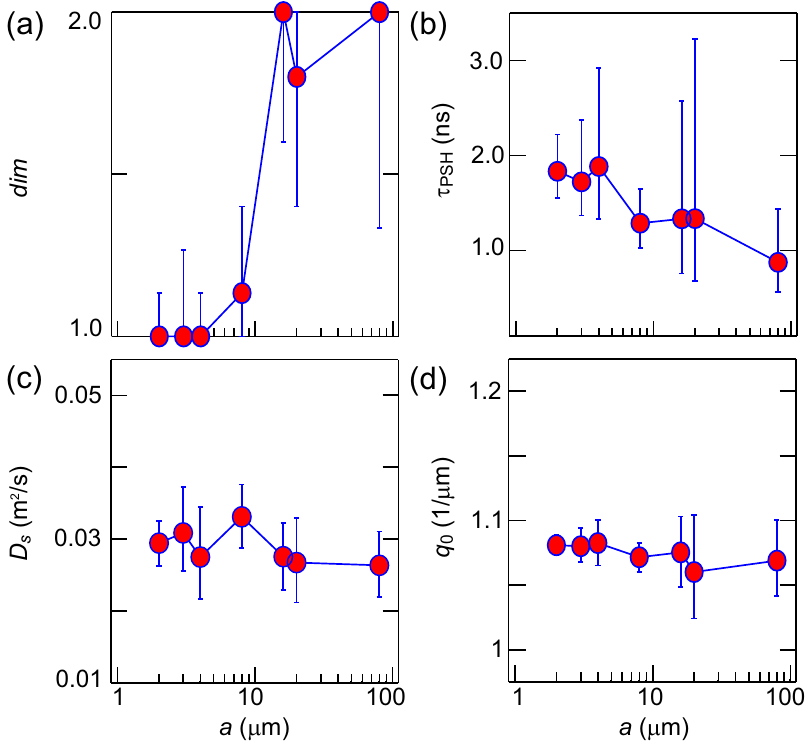}
\caption{\label{fig:fig4} Fit parameters plotted against the wire width, $a$. (a) Wires smaller than 8~$\mu$m are identified as 1D. To obtain the parameters shown in (b) to (d), $dim$ was fixed to 2 for wires wider than 8~$\mu$m and to 1 for the others. (b) The PSH lifetime, $\tau_\textrm{PSH}$, increases slightly with decreasing wire width. (c) and (d) show the spin diffusion constant $D_\textrm{s}$ and the helix wavenumber $q_0$, respectively. Both parameters do not change with $a$.}
\end{figure}

Figure~\ref{fig:fig4}(a) shows the fit parameter $dim$ obtained from Eq.~(\ref{eq1}). 
The indicated confidence interval of each parameter is defined as a maximum 5\% increase of the mean-square fit error from its global minimum when one parameter is detuned from its minimum value while all other parameters are optimized.
A clear transition to 1D diffusion is observed for wires thinner than 8~$\mu$m. Comparing with Fig.~\ref{fig:fig3}, this indeed corresponds to the group of wires with suppressed decay. The transition to 1D diffusion occurs when the width of the spin distribution becomes comparable to $a$. In the experiment presented, the spin distribution has an initial sigma width of $ w\approx$~1.5~$\mu$m (determined by fitting). Accordingly, wires $\leq$~4~$\mu$m can be expected to show 1D diffusion from the beginning, in accordance with the fit results. In addition, in the 8-$\mu$m wire, diffusive broadening leads to a transition from 2D to 1D diffusion after $t \approx [(a/2)^2 - w^2]/ 2 D_\textrm{s} \approx 0.25$~ns, making this wire predominantly 1D in our fit. The strong suppression of diffusive spin decay in 1D wires results in one order of magnitude stronger signal after 2~ns [see Fig. \ref{fig:fig3}].

For the determination of the other fit parameters [see Figs.~\ref{fig:fig4}(b)-(d)], $dim$ is fixed to  1 (2) for $a\leq 8~\mu$m ($a\geq 16~\mu$m). In addition to the strong suppression of spin decay induced by a transition to 1D diffusion, we find that $\tau_\textrm{PSH}$ slightly increases for decreasing wire width [see Fig. \ref{fig:fig4}(b)].
No influence on $q_0$ can be found, as can be seen from Fig. \ref{fig:fig4}(d), meaning that the SOI is unchanged between narrow and wide wires with $q_0 = 1.05~\mu$m$^{-1}$, corresponding to $l_{\textrm{SO}} = 6.0~\mu$m. Furthermore, also $D_\textrm{s}$ is unaffected [Fig. \ref{fig:fig4}(c)]. Hence, the electron scattering rate is not significantly enhanced by diffusive scattering at the wire edges. At the same time, a significant reduction of $n_\textrm{s}$ in thin wires can be excluded, as this would also affect $D_\textrm{s}$ and $q_0$.

The small increase of $\tau_\textrm{PSH}$ with decreasing $a$ is at first sight surprising, considering that a quadratic dependence on $a$ for $a < l_\textrm{SO}$ has been predicted for a Rashba-only model~\cite{Malshukov2000, Kiselev2000, Chang2009}. This prediction is the outcome of a correlation of the motion of spin polarization on the Bloch sphere with the spatial motion of the diffusing electrons \cite{Chang2004}. In case of such correlation, a lateral confinement restricts the spins to a ring on the Bloch sphere. Spins then mostly rotate about a fixed precession axis, leading to the formation of a helical spin mode with enhanced $\tau_\textrm{PSH}$. This applies to diffusive transport with no lower boundary for the electron mean free path. The same restriction to a ring on the Bloch sphere occurs for $\alpha = \beta$ without lateral confinement, but in a wire is expected to survive for any superposition of Rashba and linear Dresselhaus SOI.
 
The cubic Dresselhaus term $\beta_3$ can be decomposed into a contribution that is harmonic in the spherical angle $\theta$ of the electron momentum, and one that is harmonic in $3\theta$~\cite{Knap1996}. The former can be combined with the linear Dresselhaus contribution into the renormalized parameter $\beta=\beta_1-\beta_3$ and is also suppressed in a laterally confined geometry. The third harmonic component however is disturbing the correlation between spin polarization and spatial position. Because of this detrimental effect, channel confinement can not strongly suppress this decay rate in the diffusive limit. However, it is conceivable that in the quasi-ballistic regime, $a > l_\textrm{p}$, the spin decay rate due to $\beta_3$ can be decreased in case of specular scattering at the wire edges. In our sample with $l_\textrm{p} = 0.22~\mu$m~$\ll a$, this effect is not expected to play a significant role. 
In the investigated QW without wire confinement, $\tau_\textrm{PSH}$ is limited to a large extent by the third-harmonic cubic Dresselhaus contribution~\cite{Walser2012}. 
Thus, the observed small increase of $\tau_\textrm{PSH}$ is compatible with a suppression of the decay caused by the imbalance of the linear SOI, $\alpha - \beta$. But the lifetime remains limited by the cubic Dresselhaus SOI, $\beta_3$.

In this light, the strong increase of the spin lifetime for $a<l_\textrm{SO}$ observed in Refs.~\onlinecite{Denega2010, Ishihara2013} must be discussed by taking into account that the laser spot size used in these studies is larger than $l_\textrm{SO}$ and the spin diffusion length. Under this condition, the transition to 1D diffusion can not be observed as no significant amount of spins diffuses out of the area sampled by the probe beam. Furthermore, the large beam spots average out the helical spin mode, leading to a fast decay of the Kerr signal within a timescale defined by the D'yakonov-Perel mechanism~\cite{Kiselev2000, Salis2014}, which in our sample is on the order of 40~ps. No quadratic dependence of the measured spin decay rate on the wire width is then expected. However, if the wire is oriented perpendicular to the PSH direction, a large spot will see a gradually increasing lifetime for $a<l_\textrm{SO}$ because the wire edges prevent the propagation of a helical spin mode, and the probe beam only samples spin polarization of the same sign. For $a \ll l_\textrm{SO}$, the measured lifetime will approach $\tau_{\textrm{PSH}}$ even for spot sizes larger than $l_\textrm{SO}$, but only for wire directions perpendicular to the helix direction, explaining the anisotropy in the observed spin lifetime~\cite{Kiselev2000, Liu2010, Kunihashi2012}.

In conclusion, we have studied the impact of lateral confinement on the decay dynamics of a local spin excitation in a PSH system. Despite the enhanced spin dephasing time of a PSH, we find that wire confinement along the PSH direction leads to a further increase of $\tau_\textrm{PSH}$. We argue that the mechanism for this enhancement is based on a correlation between spin polarization direction and spatial position. Within this model, the effect of channel confinement depends on the particular symmetry of the SOI. Specifically, the third spherical harmonics of the cubic Dresselhaus term is not strongly affected, explaining the only weak increase of $\tau_\textrm{PSH}$.
Furthermore, a transition to 1D diffusion is observed for wires thinner than 8~$\mu$m, with a spin polarization that decays proportionally to $1/\sqrt{t}$ rather than $1/t$ as in the 2D case. This leads to a largely enhanced visibility of the PSH at longer times. After 2~ns, because of the combined benefits of 1D diffusion and suppression of imbalance terms, the spin density is enhanced by a factor of 20 in a 2-$\mu$m wire compared with an 80-$\mu$m wire. Our findings also have implications for more general situations of SOI away from the PSH case and could improve the signal not only in optical, but also in transport measurements.

Financial support from the NCCR QSIT is acknowledged.
We thank R. Allenspach, Y.~S. Chen, A. Fuhrer, M. Kohda and R.~J. Warburton for helpful discussions, and U. Drechsler and S. Reidt for technical assistance.

\end{document}